\documentclass[12pt,a4paper]{article}%

\usepackage{amsmath,amssymb,amsfonts}
\usepackage{caption2}
\usepackage[]{hyperref}
\usepackage[dvips]{graphicx}
\usepackage{multicol}

\numberwithin{equation}{section} 
\setcaptionmargin{1cm}
\newcommand{\hhref}[1]{\href{http://arxiv.org/abs/#1}{{\it arXiv:#1}}}

\begin{document}


\begin{titlepage}
$\quad$
\vskip 4.0cm
\begin{center}
{\huge \bf  Selection rules for helicity amplitudes in massive gauge theories} 
\\ 
\vskip 2.0cm {\large
{\bf Francesco Coradeschi}$^a$ and {\bf Paolo Lodone}$^a$ 
} \\[1cm]
{\it 
$^a$ Institut de Th\'eorie des Ph\'enom\`enes Physiques, EPFL, Lausanne, Switzerland 
} \\[5mm]
\vskip 1.0cm
\today
\end{center}
\begin{abstract}

After a rediscussion of the vanishing theorems for helicity amplitudes in unbroken gauge theories,  we study the case of spontaneously broken gauge theories at high energy.
The vanishing theorems generalize to a definite pattern of $m/E$-suppression of the amplitudes that vanish in the massless case, where $E$ is the energy scale of the process and $m$ is the mass of the gauge vectors.
We use only elementary arguments, and as an application we show how these methods  can be employed to understand some aspects of the effective $W$ approximation in the polarized case.

\end{abstract}
\end{titlepage}

\section{Introduction}

Scattering amplitudes play a central role in the study of gauge theories, both from the theoretical and the phenomenological side.

From the formal point of view, the properties of scattering amplitudes have often provided clues to hidden symmetries and unexpected dynamical structures in gauge theories.
For example it was realized long ago that the tree-level amplitudes are effectively supersymmetric \cite{Parke:1985pn}\cite{Kunszt:1985mg}, so that their structure is significantly constrained by the fact that they obey supersymmetric Ward identities \cite{Grisaru:1976vm}\cite{Grisaru:1977px}.
By making use of these relations, it is easy to show that the tree-level amplitude for n (massless) gauge vector scattering vanishes unless it involves at least two helicities of each sign. Schematically, one says that the helicity configurations $(\pm +...+)$ and $(\pm -...-)$ vanish while the maximally-helicity-violating (MHV) nonvanishing amplitudes are of the type $(--+...+)$ and $(++-...-)$, with all momenta ingoing.
The tree-level MHV amplitude for n-gauge boson scattering can moreover be expressed by a remarkably simple formula, which was discovered by Parke and Taylor \cite{Parke:1986gb} and later proven by Berends and Giele \cite{Berends:1987me}.
It was also found that part of this simplicity extends to the loop level, at least for N = 4 super Yang-Mills theory \cite{Bern:1994zx}, and there has been a resurgence of interest in uncovering new properties of scattering amplitudes after Witten \cite{Witten:2003nn} reformulated gauge theory as topological string theory in twistor space. Recent developments include a new type of recursion relation \cite{Britto:2004ap}\cite{Britto:2005fq} and further generalizations \cite{ArkaniHamed:2009dn}\cite{ArkaniHamed:2010kv}.

At the phenomenological level, scattering amplitudes are critical to the prediction of cross sections at high-energy colliders, for processes within and beyond the Standard Model. The efficient evaluation of amplitudes involving many quarks and gluons is particularly important at the Large Hadron Collider (LHC), in which multi-jet final states are copiously produced, see \cite{Mangano:1990by}-\cite{Peskin:2011in} for modern reviews.

The central issue at the LHC is however to test the electroweak sector, in which the gauge symmetry is spontaneously broken. A relevant question to ask is therefore how much of the simplicity of the massless case is left in the massive case, at least when the energy scale $E$ at which the system is probed is much larger than the mass scale $m$ of the gauge vectors.
Since there is no discontinuity in taking the limit $m/E \rightarrow 0$ for a gauge-invariant amplitude among given transverse-helicity states\footnote{Although the concept of helicity is not frame-independent in the massive case, as discussed below.}, we expect the tree-level helicity amplitudes that vanish in the massless case to be in general nonvanishing, but suppressed by some power of $m/E$. In particular, the MHV amplitude in the massive case is in general expected to be of the type $(+...+)$.
The purpose of this study is to understand analytically the pattern of these suppressions, focussing on the tree-level amplitudes.

In trying to extend the usual arguments it is clear that the supersymmetric one, although very elegant, is not straighforwardly generalizable to spontaneously broken gauge theories\footnote{For a generalization of the supersymmetric argument to the massive case see e.g. \cite{Boels:2011zz}.}.
An alternative way to understand why some tree-level amplitudes vanish in the massless case is to take advantage of the spinor formalism, in which the polarization vectors are expressed in terms of spinor wavefunctions.
Although it is possible to generalize the spinor formalism to the massive case\footnote{See e.g. \cite{Dittmaier:1998nn}\cite{Andreev:2001se}.}, and although this has proved to be a fruitful approach for numerical implementation, it is also clear that it is not suitable for an analytic understanding of the cancellations due to the complication of the various expressions.
We take a different approach: since the properties of the polarization vectors in the spinor formalism are related to the properties of axial gauges, we simply start from the general properties of the amplitudes in an axial gauge. Instead of making use of the spinor representation of the polarization vectors, we thus directly use the defining properties of the polarization vectors.
We show that in this way it is possible to study the $(m/E)^n$-behaviour of the helicity amplitudes in a simple and direct way. Although the concept of helicity is frame-dependent, to define the ``$+$'' and ``$-$'' states amounts to choose a basis in the space of polarizations. We show that for a particular choice of the reference vector that defines the axial gauge, namely for light-like axial gauges, the MHV amplitudes are of the type $(-+...+)$ and $(+-...-)$ while the amplitudes among equal helicities stay exactly zero. In this sense, light-like axial gauges seem to be the optimal choice in order to understand the behavior of tree-level amplitudes in the massive case.

One may argue that such a study is not phenomenologically necessary since experimental data are compared to a given model through very efficient numerical codes that do not rely on the tree-level approximation and do not need to know whether some helicity amplitudes are suppressed, since they  can compute `brute force'  and sum over all the states.
Nevertheless, in order to extract from data the physically relevant parameters and couplings it can be very useful to have an analytic although approximate control of the relevant processes.
In this sense the Effective W Approximation (EWA) originally discussed by Kunszt and Soper and recently reconsidered in \cite{Borel:2012by}, although obsolete for the true data analysis, can be a useful instrument to understand under which conditions the sensitivity to the interesting physics related to the electroweak symmetry breaking is maximized.
As an application of our study, we show that our methods can shed light on a few issues raised in \cite{Borel:2012by}.

This work is organized as follows. 
We recall some basic properties of the massless case in Section \ref{sec:axial:1} and we consider the massive case in the axial gauge in Section \ref{sec:axial:2}. In Section \ref{sec:covariant} we discuss the case of covariant gauge, in Section \ref{sec:ewa} we make some contact with the EWA, and we conclude in Section \ref{sec:conclusions}.

\section{Massless case} \label{sec:axial:1}

We start by recalling the most elementary argument, to our knowledge, that explains why the $(\pm+...+)$ and $(\pm-...-)$ amplitudes vanish in the case of unbroken gauge theories, see e.g. \cite{Dixon:1996wi}. Here and throughout the paper, we take all momenta to be ingoing.
We employ the axial gauge $k^\mu A_\mu=0$ (see Appendix \ref{app:axialgauge}), and we denote by $\epsilon_\mu^\pm(p,k)$ the polarization vector corresponding to a spin-1 particle with momentum $p$. The defining properties of such polarization vectors are:
\begin{eqnarray}
&& k\cdot \epsilon^\pm(p,k) = p\cdot \epsilon^\pm(p,k)=\epsilon^\pm (p,k) \cdot \epsilon^\pm (p,k)=0 \label{eq:axialpol:1}\\
&& \epsilon^\pm (p,k)^* = \epsilon^\mp(p,k) \\
&& \epsilon^\pm (p,k)^* \cdot \epsilon^\pm(p,k) = -1 \label{eq:axialpol:3}
\, .
\end{eqnarray}
Notice that this choice of the polarizations corresponds to the usual one of the spinor formalism in which $k$ is called the `reference vector' (see e.g. \cite{Mangano:1990by}-\cite{Peskin:2011in}), and is singular when $k\cdot p=0$.
Using either the spinor formalism, in which (\ref{eq:axialpol:1})-(\ref{eq:axialpol:3}) hold by construction, or setting $m=0$ in the arguments that are given in the subsequent Section for the massive case, one can easily show the additional properties:
\begin{eqnarray}
&& \epsilon^+ (p,k)\cdot \epsilon^+(q,k)=0 \label{eq1:massless} \\
&& \epsilon^+ (p,k) \cdot \epsilon^-(k,q)=0 \label{eq2:massless} \\
&& \epsilon^\pm_\mu (p,k) - \epsilon^\pm_\mu (p,q) \propto p_\mu \, , \label{eq3:massless}
\end{eqnarray}
where $p,q,k$ are generic light-like momenta having nonzero scalar product with each other.
Because of gauge invariance, eq. (\ref{eq3:massless}) implies that one is free to choose a different reference momentum for each polarization vector, and moreover one can choose different reference momenta in different gauge-invariant sub-amplitudes.
As shown below, these properties are enough to prove that the $(\pm+...+)$ and $(\pm-...-)$ amplitudes vanish at tree level.

Before going on, an observation is in order. In the following we will analyze the various amplitudes by counting the number of 4-vectors at our disposal at the numerator, that are the polaritazion vectors $\epsilon_\mu$ and the momenta $p^\mu$ that appear in the Feynman rules for triple vertices.
One has to notice, however, that the vector boson propagator $\mathcal{N}_{\mu\nu}(p,k)$ in axial gauge (\ref{eq:axial:propagator}) has additional terms besides the metric tensor $g_{\mu\nu}$, and thus contractions like $\epsilon_\mu \mathcal{N}^{\mu\nu}(p,k) \epsilon_\nu$ involve in general also terms of the type $\epsilon\cdot p$.
A crucial point is to notice that all these terms are not present when the reference vector is light-like, which will be the case of our interest, and that in general:
\begin{equation} \label{eq:obspropagators}
\epsilon_\mu(q_1,k) \,  \mathcal{N}^{\mu}_{\,\, \alpha_1}(p_1,k)\, \mathcal{N}^{\alpha_1}_{\,\, \alpha_2}(p_2,k) ...\, \mathcal{N}^{\alpha_{n-1}\nu}(p_n,k) \,  \epsilon_\nu(q_2,k) =
 (-1)^n \epsilon(q_1,k) \cdot \epsilon(q_2,k) + O(\frac{k^2}{E^2}) \, .
\end{equation}
where $E$ is the typical size of the components of the various momenta and of the reference vector $k$.
In this Section, the reference vector is always taken to be light-like. Of course the results do not depend on the reference vector in the case of unbroken gauge theories because of gauge invariance, but the proof is less transparent if $k^2\neq 0$ due to the additional terms in (\ref{eq:obspropagators}).

\begin{itemize}

\item 
{\bf Case $(+...+)$:} Since there are only triple and quartic vertices, the number of vertices at tree level is $\leq n-2$, where $n$ is the number of external legs and the equality holds if the vertices are all with three legs. Moreover, each vertex brings at most one momentum 4-vector (if triple, no momentum if quartic). As a consequence, each term of the amplitude must involve at least one Lorentz contraction of two polarization vectors. Since the helicities are all positive, all such contractions vanish in axial gauge thanks to (\ref{eq1:massless}).

\item 
{\bf Case $(-+...+)$:} As above, each term of the amplitude involves at least one Lorentz contraction of two polarization vectors. All the terms $\epsilon^+(p_i,k) \cdot \epsilon^+(p_j,k)$ vanish again provided that the reference vector $k$ is the same for all the positive helicities. We are then left with terms of the type  $\epsilon^+(p_i,k) \cdot \epsilon^-(p_1,k_1)$, that vanish by (\ref{eq2:massless}) if we choose $k=p_1$. Of course the reference vector of $\epsilon^-$ must be a  different 4-vector $k_1\neq k$, but this can always be done thanks to (\ref{eq3:massless}), as already observed.

\end{itemize}

The amplitudes $(-...-)$ and $(+-...-)$ also vanish in full analogy, while there is no suppression in general in the amplitudes involving at least two polarizations of each type.
Other useful results concern the amplitudes involving vector bosons plus two massless scalar or fermion legs, which also vanish when all the vectors carry the same helicity.
In the literature, this is usually shown using the supersymmetric Ward identities (see e.g. \cite{Mangano:1990by}\cite{Dixon:1996wi}).
We present here alternative and more elementary arguments that only make use of the properties of the polarization vectors \eqref{eq:axialpol:3}-\eqref{eq3:massless}, along the same lines as the discussion above.

\begin{itemize}

\item 
{\bf Case $(\phi\phi^* +...+)$:} In the case of an amplitude with $n$ vectors and two scalar external legs, the number of vertices at tree level is $\leq n$, and the equality holds when all the vertices are triple. 
We thus need to consider only the amplitudes in which all gauge vertices are triple, since the others give a vanishing contribution because of (\ref{eq1:massless}).

\noindent Let us call $p_1$ the momentum of one scalar external leg, and let us consider the sub-piece of the amplitude corresponding to its first scalar-vector branching, with the emitted vector eventually ending in $j\leq n$ vectors with momenta $\{  k_1, ... k_j  \}$. 
Taking all the reference vectors to be equal to $p_1$, this subamplitude is of the form:
\begin{equation} \label{eq:interm:step1}
(2 p_1+ \sum_{i=1}^j k_i)^{\mu} \, \mathcal{A}_{\mu} (k_1, ... k_j, \epsilon(k_1,p_1), ... \epsilon(k_j,p_1)) \, 
\end{equation}
where $\mathcal{A}_{\mu}$ is the subamplitude attached to the vector line.
We now show inductively that this expression vanishes\footnote{Notice that this can be shown more formally using BRST invariance, in particular the fact that a longitudinally polarized gauge boson is the BRST transformation of a ghost field.}.  For $j=1$, this is true because $\mathcal{A}_{\mu} (k_1, \epsilon(k_1,p_1)) \equiv \epsilon_\mu(k_1,p_1)$ which is orthogonal both to $p_1$ and to $k_1$ by definition (\ref{eq:axialpol:1}).
For $j>1$, let us consider the first splitting of our vector line into two vector lines (we know that there are only triple vertices).
To conclude it is sufficient to show that in all the terms in which (\ref{eq:interm:step1}) splits there is a factor in which the subamplitude attached to a vector line is again contracted only with the reference vector $p_1$ or with the momentum along the line. This is sufficient because eventually we will get to an external leg and the contraction with the polarization vector will give zero as in the case $j=1$.

\noindent Call in fact $p = -\sum_{i=1}^j k_i$, $\, q = \sum_{i=1}^r k_i$ and $k=\sum_{i=r+1}^j k_i$ the momenta ingoing in the triple gauge vertex.
The expression in (\ref{eq:interm:step1}) is proportional to:
\begin{equation} \label{eq:interm:step2}
(2 p_1-p)^{\mu} \, \frac{\mathcal{N}_{\mu\nu} (p,p_1)}{p^2} \,
V(p,q,k)^{\nu\rho\sigma} \,
\mathcal{A}_{\rho}(q) \,
\mathcal{A}_{\sigma}(k)
\end{equation}
where we denoted by $\mathcal{A}_{\rho}(q)$ and $\mathcal{A}_{\rho}(k)$ the subamplitudes attached to the new vector lines, and (see Appendix \ref{app:axialgauge}):
\begin{eqnarray}
\mathcal{N}_{\mu\nu} (p,p_1) &=& -g_{\mu\nu} + \frac{p_\mu (p_1)_\nu + p_\nu (p_1)_\mu}{p\cdot p_1} - \frac{(p_1)^2}{(p\cdot p_1)^2} p_\mu p_\nu \label{subeq:1}
\\
V(p,q,k)^{\nu\rho\sigma} &=& g^{\sigma \nu}(k-p)^\rho + g^{\nu\rho}(p-q)^\sigma + g^{\rho\sigma}(q-k)^\nu \, . \label{subeq:2}
\end{eqnarray}
By expanding (\ref{eq:interm:step2}) using (\ref{subeq:1}) and (\ref{subeq:2}) it is easy to see that in all the various terms either $\mathcal{A}_{\rho}(q)$ or $\mathcal{A}_{\sigma}(k)$ is contracted either with the corresponding momentum ($q$ or $k$, respectively) or with the reference vector $p_1$. This is true apart from the term involving $g^{\rho\sigma}$, but this gives no contribution because it necessarily involves also the contraction of two polarization vectors.

\noindent As a consequence, these amplitudes vanish at tree level.

\item 
{\bf Case $(\psi\psi^* +...+)$:}  
 Consider first of all the fermion wavefunctions that appear in the amplitude. With massless fermions, there are only two independent solutions of the Dirac equation,\footnote{Recall that there is no difference between $u(p)$ and $v(p)$ up to normalization conventions. This has to be the case since the projectors $ \sum u(p)\overline{u}(p) $ and $\sum v(p)\overline{v}(p) $ are both equal to $ {\not p}$.} that can be written down as:
\begin{equation}
\psi_{\pm}(p) = \frac{1}{2}(1 \pm \gamma^5) u(p)
\quad , \quad
{\not p} \, u(p) = {\not p} \, \psi_{\pm}(p) =  0\, .
\end{equation}
By definition these wavefunctions are eigenvectors of rotations along the axis specified by $\vec{p}$, under which they transform as:
\begin{equation}
\psi_{\pm}(p) \rightarrow R^{\vec{p}}_{\theta} \psi_{\pm}(p) = e^{\mp i\frac{\theta}{2}} \psi_{\pm}(R^{\vec{p}}_{\theta}p) = e^{\mp i\frac{\theta}{2}} \psi_{\pm}(p)  \, .
\end{equation}
At the same time, as shown in eq. (\ref{eq:trasfpol:transv2}), the transformation properties of $\epsilon^{\pm}(k,p)$ under the same rotation are:
\begin{equation}
\epsilon^{\pm}(k,p) \rightarrow R^{\vec{p}}_{\theta} \epsilon^{\pm}(k,p)= e^{\pm i{\theta}}\epsilon^{\pm}(R^{\vec{p}}_{\theta}k,R^{\vec{p}}_{\theta} p)
= e^{\pm i{\theta}}\epsilon^{\pm}(R^{\vec{p}}_{\theta}k,p) \, .
\end{equation}
On the other hand, since $\{{\not p}, {\not  \epsilon}^{\pm}(k,p)\} =0$, the four spinors ${\not \epsilon}^{+,-}(k,p) \psi_{+,-}(p)$ satisfy the Dirac equation so that each of them must be a linear combination of $\psi_+(p)$ and $\psi_-(p)$.
But the above transformation properties under rotations along $\vec{p}$ can match only in two cases\footnote{One can further show that ${\not \epsilon^{\pm}(k,p)} \propto {\psi}_{\pm}(p) \overline{\psi}_{\pm}(k)  +  {\psi}_{\mp}(k) \overline{\psi}_{\mp}(p)$, from which (\ref{eq:fermions:mainpoint}) follows, see e.g. \cite{Mangano:1990by}. \label{foot:spinors}}, and thus the other two must vanish:
\begin{eqnarray}
{\not \epsilon}^{+} (k,p) \psi_- (p) &=&  \overline{\psi}_+ (p) {\not \epsilon}^{+} (k,p) =0 \, ,
 \label{eq:fermions:mainpoint} \\
{\not \epsilon}^{-} (k,p) \psi_+ (p) &=&  \overline{\psi}_- (p) {\not \epsilon}^{-} (k,p) =0 \nonumber
\end{eqnarray}

\noindent For the amplitude with $n$ vectors and two massless-fermion legs with momenta $p_1$ and $p_2$, the number of vertices at tree level is again $\leq n$, and again the equality holds when all the gauge vertices are triple.
Moreover because of (\ref{eq1:massless}) we know that, when all the vectors have the same helicity, we can restrict ourselves to the terms in which all the gamma matrices are Lorentz-contracted with a polarization vector and never with a momentum 4-vector, otherwise a term $\epsilon \cdot \epsilon$ must be present.
To conclude it is enough to notice that the helicity of the fermion line is necessarily conserved, that is, only the cases $\overline{\psi}_+(p_1){\not \epsilon} ... {\not \epsilon}\psi_+(p_2)$ and $\overline{\psi}_-(p_1){\not \epsilon} ... {\not \epsilon} \psi_-(p_2)$ can be nonvanishing.
Using (\ref{eq:fermions:mainpoint}) we then see that all the terms in the amplitude vanish if we choose as reference vector $p_1$ in the case of right-handed fermions, or $p_2$ in the case of left-handed fermions.

\end{itemize}

In conclusion, in this Section we reviewed known results about the vanishing of helicity amplitudes at tree level in an unbroken gauge theory. In doing so, we made use only of basic properties of the polarization vectors in the axial gauge together with other elementary considerations.
As is well known, these results greatly simplify the computation of tree-level scattering amplitudes in massless gauge theories, making very convenient for practical computations to use the spinor formalism with light-like reference vectors, in which the polarization vectors have the same properties as in the axial gauge with light-like reference vector.
%
Notice also that in the case $(+...+)$, in order to see the cancellation, it is simply enough to employ an axial gauge while in the other cases it is crucial to choose properly the reference vector(s).
As discussed in Section \ref{sec:axial:2}, this is the basic reason why in the massive case the $(+...+)$ amplitude is suppressed by more powers of $m/E$ than the others.

\section{Massive case} \label{sec:axial:2}

In order to understand the high-energy behaviour of the above helicity amplitudes in a broken gauge theory, the first step is to generalize (\ref{eq1:massless})-(\ref{eq3:massless}) to the massive case.
We consider the case in which the typical center-of-mass (CM) energy $E$ of the process is much larger than the mass $m$ of the vector boson, i.e. $\varepsilon \equiv m/E \ll 1$.
For simplicity, we consider the case in which the mass $m$ is the same for all the gauge vectors.
Generally speaking, a tree level amplitude involving $n$ external legs has mass dimension $4-n$, so that we can write, working in the CM frame:
\begin{equation} \label{eq:suppr:general}
\mathcal{A}_{(n)} \sim E^{4-n} \varepsilon^{t} \, .
\end{equation}
Since no gauge-invariant tree-level amplitude should become singular in the limit $m\rightarrow 0$, we expect $t\geq 0$. 
Notice that this is true in a ``physical'' gauge like the axial one, and thus it must be true in general for all the gauge-invariant subamplitudes. On the contrary non-gauge-invariant subamplitudes can grow faster with energy, and this behaviour cancels out only in the gauge-invariant results.
As discussed in Section \ref{sec:covariant}, this is what happens in covariant gauges in which the longitudinal polarization vectors scale as $E/m$ introducing a ``bad high-energy behaviour'' that cancels away only in the gauge-invariant expressions.
For this reason the axial gauge is the most suitable for an analytic understanding of the high-energy behaviour of the helicity amplitudes in the massive case, and moreover it might be useful from the computational point of view since it does not involve large cancellations among subamplitudes. 

We also expect the polarized amplitudes that vanish in the massless case to be suppressed  in (\ref{eq:suppr:general}) by some exponent $t>0$, that we want to compute, or to stay equal to zero.
In order to do that, as noted in \cite{Borel:2012by}, an useful selection rule can be obtained by noticing that the Lagrangian of the Goldstones $s$ and gauge fields is invariant under the reparametrization:
\begin{equation} \label{eq:mtominusm}
A_\mu \rightarrow A_\mu
\quad , \quad
s \rightarrow -s
\quad , \quad
m \rightarrow -m
\, .
\end{equation}
As a consequence, the sign of $m$ is not a physical observable and any amplitude among given external states must be either even or odd under $m \rightarrow -m$, since iterating (\ref{eq:mtominusm}) twice gives no change,  and thus $t$ must be an integer in (\ref{eq:suppr:general}).
Moreover we see from (\ref{eq:axial:longitud}) that the longitudinal polarization vector changes its sign under (\ref{eq:mtominusm}), while the transverse ones are unchanged, and thus we can transform an even amplitude into an odd one and vice-versa by changing a transverse external state into a longitudinal one. Since the amplitudes involving only transverse states are even under (\ref{eq:mtominusm}), this implies that the polarized amplitudes must be even or odd under $m \rightarrow -m$ depending on whether they involve an even or an odd number of longitudinal states.

Another useful argument is the one of recovering the massless case in the limit $m\rightarrow 0$.
In fact in this limit the theory approaches an unbroken gauge theory interacting with massless scalars, for which the results of the Section \ref{sec:axial:1} hold.
In particular, we deduce that all the amplitudes among transverse states that vanish in the massless case must have $t>0$ in (\ref{eq:suppr:general}). Moreover, when we take this limit in amplitudes involving longitudinal states, we can trade the longitudinally-polarized vectors for massless scalars charged under the gauge group\footnote{This is also evident if we look at the explicit form of the longitudinal polarization vectors (\ref{eq:axial:longitud}), that makes the equivalence theorem an identity.}.

Notice however that, differently from the massless case, our polarizations $\epsilon^+(p,k)$ and $\epsilon^-(p,k)$ do not describe the helicity eigenstates in a Lorentz-invariant way since the helicity itself is not a Lorentz-invariant concept.
The physical states described by the polarization vectors $\epsilon^\pm(p,k)$ are elements of a basis in the space of polarizations, and in the massive case this basis depends crucially on the vector $k^\mu$. For this reason, in order for a polarized amplitude to have a physical meaning, it is now necessary that the reference vector be the same for all the polarizations.
Analogously, as discussed  in Section \ref{sec:covariant}, the physical state described by $\epsilon^+(p,k)$ is in general not the same as that described by the usual `plus' polarization in a covariant gauge $\epsilon^+(p)$, defined below.
As we will see, in the massive case the basis of ``axial-gauge-like'' polarizations  with light-like reference vector are a particularly convenient choice.

To generalize (\ref{eq1:massless})-(\ref{eq3:massless}), we start from the projector on the transverse subspace. Using the full projector (\ref{eq:axial:propagator}) and the explicit form of the 4-vector part of the longitudinal polarization (\ref{eq:axial:longitud}),
we can write the transverse projector as:
\begin{eqnarray}
\sum_{\lambda = \pm} \epsilon^\lambda_\mu(p,k) \epsilon^\lambda_\nu(p,k)^* &=& 
\epsilon^+_\mu(p,k) \epsilon^-_\nu(p,k) + \epsilon^-_\mu(p,k) \epsilon^+_\nu(p,k) \label{eq:projector} \\
&=& -g_{\mu\nu} - \frac{1}{(pk)^2 - p^2 k^2} \left( p^2 k_\mu k_\nu + k^2 p_\mu p_\nu - pk(p_\mu k_\nu + k_\mu p_\nu) \right),
\nonumber
\end{eqnarray}
with $p^2=m^2$ and for any value of $k^2$. 
To generalize (\ref{eq1:massless}), we contract (\ref{eq:projector}) with $\epsilon^+_\mu(q,k)$ and square the two sides of the resulting equation. We obtain:
\begin{equation}
2 [\epsilon^+(p,k)\cdot \epsilon^+(q,k)] \, [\epsilon^-(p,k)\cdot \epsilon^+(q,k)] =  - k^2 \frac{[p\cdot \epsilon^+(q,k)]^2}{(pk)^2 - p^2 k^2} \, ,
\end{equation}
where it is always understood that the momenta of the vectors are on shell, that is $p^2 = q^2 = m^2$.
By continuity with the case $q \to p$, there can not be suppression in $\epsilon^-(p,k)\cdot \epsilon^+(q,k)$. We conclude that:
\begin{equation}
\epsilon^+(p,k)\cdot \epsilon^+(q,k) = O(\frac{k^2}{E^2}) \label{eq1:massive}
\end{equation}
where $E\gg m$ is the typical size of the entries of $p_\mu$ and $k_\mu$ in the CM frame. Notice that the size of the entries of $k^\mu$ does not have any physical meaning, since the projector \eqref{eq:axial:propagator} is invariant under a rescaling of the reference vector by an arbitrary factor. For notational convenience, we choose the leading nonzero entries of $k^\mu$ to be of order $E$.

To generalize (\ref{eq2:massless}) we do the same as above using $\epsilon^-_\mu(k,q)$, and we find:
\begin{eqnarray}
\epsilon^+(p,k)\cdot \epsilon^-(k,q) &=& O(\frac{k^2}{E^2}) \label{eq2:massive}\\
\epsilon^+(p,k)\cdot \epsilon^-(k,p) &=& 0 \, . \label{eq2:massiveBis}
\end{eqnarray}
We also need to generalize (\ref{eq3:massless}). This can be done by writing down the explicit expression for the polarization vectors and expanding it for small $p^2/E^2=m^2/E^2$. The result is:
\begin{equation}
\epsilon^\pm_\mu (p,k) - \epsilon^\pm_\mu (p,q) =  O(1) \frac{p_\mu}{E} + O(\frac{p^2}{E^2})\frac{q_\mu}{E} + O(\frac{p^2}{E^2})\frac{k_\mu}{E} \, .
\label{eq3:massive}
\end{equation}

A final observation before going on.
As in the case $m=0$, a crucial point in the various arguments for the cancellation or suppression of the amplitudes is the analysis of the possible contractions of the polarization vectors. To this end, note that massive gauge theories in axial gauge provide an additional possible source of four-momenta which can contract with the polarizations: the scalar-vector propagators $\mathcal{N}_{\mu s}$. Using eq. \eqref{eq:axial:propagator}, it follows immediately:
\begin{equation}
\epsilon_\mu^a (p,k) \, \mathcal{N}^{\mu s} (q,k) = O(\frac{m\, k^2}{E^3}).
\end{equation}
We are now ready to extend the discussion of Section \ref{sec:axial:1} to the massive case.

\begin{itemize}

\item 
{\bf Case $(++...+)$:} Again the number of vertices at tree level is $\leq n-2$, where $n$ is the number of external legs. It is also still true that each vertex brings at most one four momentum, and thus each term of the amplitude must be proportional to the Lorentz contraction of two polarization vectors, keeping (\ref{eq:obspropagators}) in mind.
Since the helicities are all positive, all such contractions are suppressed at the level of $O(k^2/E^2)$ by (\ref{eq1:massive}) and (\ref{eq:obspropagators}). But in this case there is no condition on $k^2$, in particular we can choose $k^2=0$.
This argument shows that this kind of amplitudes vanishes at tree level in the massive case as well, provided that one chooses a light-like reference vector.

\item 
{\bf Case $(-+...+)$:} For continuity with the case $m=0$, we already know that this amplitude is suppressed by $\varepsilon^t$ with $t>0$, and from the selection rule discussed above we know that $t$ is an even integer and thus $t\geq 2$. 
Through an explicit computation one can then see that the suppression is indeed no more than this, i.e. $t=2$.

An intuitive way to understand why $t=2$ can be to recall that, in the proof of the vanishing theorem for the case $(-+...+)$ in the previous Section, we had to make a precise choice of reference vectors. In particular the reference vectors for the positive-helicity polarizations had to be all equal to the momentum of the negative-helicity leg. Since this momentum is now time-like, we expect the suppression to be no more than $t=2$ because of (\ref{eq:obspropagators}).

\item 
{\bf Case $(0+...+)$:} From (\ref{eq:axial:longitud}) we see that, with $k^2=0$, we have $\epsilon^0(p,k) \propto k$ and thus all the products $\epsilon^0(p_1,k) \cdot \epsilon^+(p_i,k)$ vanish.
As recalled in the Appendix \ref{app:axialgauge} the longitudinal polarization has also a fifth `scalar' component, but when this is involved there must be either a contraction $\epsilon^+(p_i,k) \cdot \epsilon^+(p_j,k)$ or $\epsilon_\mu^+ (p_i,k) \, \mathcal{N}^{\mu s} (q,k)$ ($q$ being some linear superposition of the $p_j$), both of which vanish for $k^2=0$. This shows that these amplitudes vanish at tree level in the massive case as well, provided that one chooses a light-like reference vector.

\item 
{\bf Case $(0-+...+)$:} As above, choosing $k^2=0$ we have $\epsilon^0(p,k) \propto k$ and all the $\epsilon^+(p_i,k) \cdot \epsilon^+(p_j,k) $ vanish. There is however no suppression in the contractions $\epsilon^-(p_2,k) \cdot \epsilon^+(p_i,k) $. In this case, taking the 4-vector part of the longitudinal polarization, we have an $O(\varepsilon)$ suppression because the components of $\epsilon^0_\mu$ are of $O(m/E)$.
The same suppression holds for the entire amplitude due to the selection rule (\ref{eq:mtominusm}) discussed above.
In conclusion, these amplitudes are suppressed by a factor $\varepsilon$.

\item 
{\bf Case $(00+...+)$:} This amplitude is even under (\ref{eq:mtominusm}), and moreover the amplitude involving only positive polarizations and two scalar lines vanishes in the massless case, as discussed in Section \ref{sec:axial:1}. We deduce that $t$ is even and $\geq 2$ in (\ref{eq:suppr:general}).
To convince ourselves that it is actually $t=2$, one can notice that the proof in the massless case relied on a specific choice of the reference vector, namely the momentum of a scalar line, and that there was no freedom to change it apart from choosing the momentum of the other scalar line. Analogously to the case $(-+...+)$ discussed above, one can then see that these amplitudes must scale as $\varepsilon^2$.

\item 
{\bf Case $(\psi {\psi}^* +...+)$:} Going through the proof that these amplitudes vanish at tree level in the massless case, we see that it is valid also for a spontaneously broken gauge theory provided that the fermion is massless and we choose a light-like reference vector.
If the fermion has a mass $m_\psi$, we expect these amplitudes to be suppressed by only one power of $m_\psi/E$.

\end{itemize}

These findings are summarized in Table \ref{table:axial}.
Notice that this choice of gauge is very convenient for tree level computations since the amplitudes $(+...+)$, $(0+...+)$ and $(\psi {\psi}^* +...+)$ all vanish (the last one only if $m_\psi=0$).
Moreover  each polarized amplitude is given by a sum of terms that are individually not parametrically larger than the total, i.e. there are never large cancellations among terms that individually grow faster than the total when the energy increases.

\begin{table}[t]
\begin{center}
\begin{tabular}{rrc}
	\multicolumn{3}{c}{$\epsilon_3 , ... \epsilon_n = +, ...+$}	\\
	\cline{1-3}
	$\epsilon_1$	&$\epsilon_2$ &Scaling 					\\
	$+$				&$+$			&$0$					\\
	$-$				&$+$			&$\varepsilon^2$					\\
	$-$				&$-$			&$\varepsilon^0$		\\
	$0$				&$-$			&$\varepsilon$					\\
	$0$				&$0$			&$\varepsilon^2$
\end{tabular}
\begin{tabular}{rrc}
	\multicolumn{3}{c}{$\epsilon_3, \epsilon_4 , ... \epsilon_n = 0,+, ...+$}	\\
	\cline{1-3}
	$\epsilon_1$	&$\epsilon_2$ &Scaling 					\\
	$+$				&$+$			&$0$					\\
	$-$				&$+$			&$\varepsilon$					\\
	$-$				&$-$			&$\varepsilon$					\\
	$0$				&$-$			&$\varepsilon^0$					\\
	$0$				&$0$			&$\varepsilon$
\end{tabular}
\end{center}
\caption{\footnotesize{Scaling of the tree-level polarized amplitudes $(\epsilon_1,...\epsilon_n)$, in the axial gauge, with the parameter $\varepsilon=m/E\ll 1$, see text. All the momenta are ingoing. The same suppression holds for the amplitudes related to the ones above by $(+ \leftrightarrow -)$ in all the external states.}}
\label{table:axial}
\end{table}

\section{Polarized amplitudes in covariant gauge} \label{sec:covariant}

In order to make contact with the more widely used covariant (Landau) gauge with $\partial_\mu A^\mu =0$, that implies $\epsilon(p)\cdot p=0$, we notice that the simplifications that occur in the axial gauge are crucially due to the structure (\ref{eq:axialpol:1})-(\ref{eq:axialpol:3}) of the polarization vectors.
To understand the pattern of $m/E$ suppressions in the polarized amplitudes in covariant gauge, it is then convenient to pass through a basis that resembles the axial one as much as possible.

Recall that, for a generic on-shell 4-momentum:
\begin{equation}
p_\mu = (E_p \equiv \sqrt{m^2 + p^2}, \ \vec{p} \ ),
\end{equation}
the polarization vectors corresponding to the three helicity eigenstates can be written as:
\begin{equation} \label{helicity}
 \epsilon^{\pm}_\mu(p) = \frac{1}{\sqrt{2}} (0, \ \hat{a}_p \pm i \, \hat{b}_p \ )
\quad , \quad
 \epsilon^{0}_\mu(p) = \frac{1}{m} (|\vec{p}\,|, E_p \hat{p} ),
\end{equation}
where $\hat{p} \equiv \vec{p}/|\vec{p}\,|$ and $\hat{a}_p$, $\hat{b}_p$ are two three-vectors with unit norm, chosen in such a way that $(\hat{a}_p,\, \hat{b}_p,\, \hat{p})$ form a right-handed orthonormal basis of the space-like $\mathbb{R}^3$. Of course, $\hat{a}_p$ and $\hat{b}_p$ are only defined up to a rotation along the $\hat{p}$ axis, and a different choice of $\hat{a}_p$, $\hat{b}_p$ leads to the appearance of an overall multiplicative phase in $\epsilon^{\pm}(p)$.

As already said, in the context of massive vector bosons, helicity means ``spin along the $\vec{p}$ direction'' and it is well-defined when $\vec{p} \neq 0$ although it is not a Lorentz-invariant concept. It is always implied  that the helicities we consider are defined in a given reference frame, and we are particularly interested in the case of the CM frame.
In fact the above definitions \eqref{helicity} are not covariant, and a generic Lorentz transformation mixes the different helicities. The helicity states are however not mixed by rotations, that is:
\begin{equation} \label{eq:rotationOfPolariz}
R_\mu^{ \nu} \epsilon^a(p)_\nu \propto \epsilon^a(Rp)_\mu, \quad a=\pm, 0
\end{equation}
where $R$ is a generic rotation, $Rp$ is the rotated momentum, and we use the symbol $\propto$ to indicate the possible appearance of phase factors (see also Appendix \ref{app:axialgauge}). As for boosts, any boost along the $\vec{p}$ axis leaves $\epsilon^0(p)$ invariant, and it can either interchange or leave invariant the $\epsilon^{\pm}(p)$ helicity states, depending on whether the sign of $\vec{p}$ is reversed or not. Finally, a boost not aligned with $\vec{p}$ mixes all the three helicity states.

Now, let $k$ be a generic 4-momentum such that $p\cdot k\neq 0$ and $\hat{k}\neq \hat{p}$; we want to construct two linear superpositions of the helicity eigenstates \eqref{helicity} that satisfy the conditions (\ref{eq:axialpol:1})-(\ref{eq:axialpol:3}). Since we are free to rotate the three-vectors $\hat{a}_p$ and $\hat{b}_p$, we can choose them such that $\hat{b}_p \cdot \vec{k} =0$.
Then it is easy to check that the 4-vectors that we look for can be written as:
\begin{equation}
\label{explpm}
\begin{split}
\hat{\epsilon}^{\pm}_\mu (p,k) =  & \frac{1}{2} \left( \frac{ \sqrt{(k \cdot p)^2 - (k_\perp^2 + k^2) m^2}}{\sqrt{(k\cdot p)^2 - k^2 m^2}}  \pm1  \right) \epsilon^{+}_\mu (p) + \frac{1}{\sqrt{2}} \frac{ k_\perp m}{\sqrt{(k\cdot p)^2 - k^2 m^2}} \ \epsilon^0_\mu (p) \, +\\
& \frac{1}{2} \left( \frac{ \sqrt{(k \cdot p)^2 - (k_\perp^2 + k^2) m^2}}{\sqrt{(k\cdot p)^2 - k^2 m^2}}  \mp1  \right) \epsilon^{-}_\mu (p),
\end{split}
\end{equation}
where $k_\perp$ is the norm of the tranverse component of $\vec{k}$ with respect to $\vec{p}$:
\begin{equation}
k_\perp = \left| \vec{k} \right|  \sqrt{1-\left(\hat{p} \cdot \hat{k} \right)^2}
\quad , \quad
\hat{k} \equiv \frac{\vec{k}}{|\vec{k}|}
\, .
\end{equation}
Notice that the second term in the right-hand side of eq. (\ref{explpm}) becomes proportional to $p_\mu$ in the limit $m\rightarrow 0$, and that one of the two other terms vanishes while the other reduces to $\epsilon^{\pm}_\mu(p)$. 
In this limit (\ref{explpm}) simply expresses the fact that the polarization vectors $\hat{\epsilon}^{\pm}(p,k)$ and ${\epsilon}^{\pm}(p)$ represent the same physical state.
The ``axial-covariant'' polarizations denoted by $\hat{\epsilon}^{\pm}(p,k)$ can be thus defined in the standard covariant gauges as linearly-independent superpositions of the standard polarization vectors ${\epsilon}^{\pm,0}(p)$. 
As such, they can be completed to a full set of polarizations by defining a third vector orthogonal to the first two and to $p$, that we denote by $\hat{\epsilon}^0(p,k)$.

The crucial observation is now that a given polarized amplitude is gauge-invariant and it must depend only on the external states, that are specified by the definition of the polarization vectors.
In particular, the relations (\ref{eq:axialpol:1})-(\ref{eq:axialpol:3}) are enough to fully specify which external states correspond to what we denote as ``$+$'' and ``$-$'' polarizations, and thus also the longitudinal one is fixed.
As a consequence, a polarized amplitude computed in covariant gauge with the polarizations defined as in (\ref{explpm}) must coincide with the corresponding amplitude computed in the axial gauge with the same reference vector.
Notice that this equality is nontrivial since the Feynman rules are different and so is $\hat{\epsilon}^0(p,k)\neq {\epsilon}^0(p,k)$, as the first one is orthogonal to $p$, the second one to $k$. Since $\hat{\epsilon}^0(p,k)$ scales as $E/m$ at high energy, we expect that large cancellations among subamplitudes are now present at least in the polarized amplitudes involving longitudinal polarizations, although the final result is the same due to the fact that gauge invariance is broken only spontaneously.

To be specific, let us check these considerations in the phenomenologically interesting case of $W-W$ scattering\footnote{We consider the $SU(2)_W$ interactions only, that is we set to zero the hypercharge and we do not include the QCD interactions}.
We compute the scattering amplitudes using FeynArts/FormCalc \cite{FeynArts1}\cite{FeynArts2}, and we show in Table \ref{table:covariant} the results for the various polarized amplitudes for the $2\rightarrow 2$ scattering using the polarization vectors (\ref{explpm}).
Taking into account that $\epsilon^\pm \rightarrow \epsilon^\mp$ when an incoming external leg becomes outgoing\footnote{In doing so one also changes the sign of the external momentum, so that an analytic continuation is understood in order to have positive energies. However one can also notice that the arguments of Sections \ref{sec:axial:1} and \ref{sec:axial:2} do not depend on the fact that the momenta are ingoing, and an $\epsilon^+$ can always be considered as an $(\epsilon^-)^*$.} and vice-versa, we see that the pattern of $\varepsilon^t$ suppressions is exactly the same as the one that was deduced in the axial gauge in Table \ref{table:axial}.

The only exception is $(0,+;+,0)\sim \varepsilon^2$, while $(0,+;0,+)\sim \varepsilon^0$. This is due to the conservation of the electric charge, that forbids some of the diagrams. Generally speaking, the suppression pattern of Table \ref{table:axial} is to be regarded as the ``minimum amount of suppression'', and it is always possible to have additional suppressions because of other reasons.
In this case it is easy to understand why this amplitude is suppressed, using the methods of Sections \ref{sec:axial:1} and \ref{sec:axial:2}.
In fact in the configuration $(0,+;+,0)$ there is a change of electric charge in the scalar line, and then the only diagrams that contribute at tree level are those with two $s-s-v$ vertices (where $s$ means scalar and $v$ means vector), and no $v-v-v$ nor $s-s-v-v$ ones. As a consequence this amplitude vanishes if we employ the momentum of one of the scalars as reference vector, as in the discussion of the $(\phi\phi^*+...+)$ case in Section \ref{sec:axial:1}. By using the arguments of Section \ref{sec:axial:2} (the rule (\ref{eq:mtominusm}) and the continuity with the massless case), it is clear that the suppression becomes at least $\varepsilon^2$ when we use a light-like reference vector.

\begin{table}[t]
\begin{center}
\begin{tabular}{rrc}
	\multicolumn{3}{c}{$W_+^{\textrm{out}},\,W_-^{\textrm{out}}=+\,+$}	\\
	\cline{1-3}
	$W_+^{\textrm{in}}$	&$W_-^{\textrm{in}}$ &Scaling 					\\
	$+$				&$+$			&$\varepsilon^0$					\\
	$0$				&$+$			&$\varepsilon$					\\
	$-$				&$+$			&$\varepsilon^2$					\\
	$+$				&$0$			&$\varepsilon$					\\
	$0$				&$0$			&$\varepsilon^2$					\\
	$-$				&$0$			&$0$					\\
	$+$				&$-$			&$\varepsilon^2$					\\
	$0$				&$-$			&$0$					\\
	$-$				&$-$			&$0$
\end{tabular}
\hspace{-3pt}
\begin{tabular}{rrc}
	\multicolumn{3}{c}{$ W_+^{\textrm{out}},\,W_-^{\textrm{out}}= 0\,0 $}	\\
	\cline{1-3}
	$W_+^{\textrm{in}}$	&$W_-^{\textrm{in}}$ &Scaling 					\\
	$+$				&$+$			&$\varepsilon^2$					\\
	$0$				&$+$			&$\varepsilon$					\\
	$-$				&$+$			&$\varepsilon^0$					\\
	$+$				&$0$			&$\varepsilon$					\\
	$0$				&$0$			&$\varepsilon^0$					\\
	$-$				&$0$			&$\varepsilon$					\\
	$+$				&$-$			&$\varepsilon^0$					\\
	$0$				&$-$			&$\varepsilon$					\\
	$-$				&$-$			&$\varepsilon^2$
\end{tabular}
\hspace{-3pt}
\begin{tabular}{rrc}
	\multicolumn{3}{c}{$W_+^{\textrm{out}},\,W_-^{\textrm{out}}=+,\,-$}	\\
	\cline{1-3}
	$W_+^{\textrm{in}}$	&$W_-^{\textrm{in}}$ &Scaling 					\\
	$+$				&$+$			&$\varepsilon^2$					\\
	$0$				&$+$			&$\varepsilon$					\\
	$-$				&$+$			&$\varepsilon^0$					\\
	$+$				&$0$			&$\varepsilon$					\\
	$0$				&$0$			&$\varepsilon^0$					\\
	$-$				&$0$			&$\varepsilon$					\\
	$+$				&$-$			&$\varepsilon^0$					\\
	$0$				&$-$			&$\varepsilon$					\\
	$-$				&$-$			&$\varepsilon^2$
\end{tabular}
\hspace{-3pt}
\begin{tabular}{rrc}
	\multicolumn{3}{c}{$W_+^{\textrm{out}},\,W_-^{\textrm{out}}=0\,+$}	\\
	\cline{1-3}
	$W_+^{\textrm{in}}$	&$W_-^{\textrm{in}}$ &Scaling 					\\
	$+$				&$+$			&$\varepsilon$					\\
	$0$				&$+$			&$\varepsilon^0$					\\
	$-$				&$+$			&$\varepsilon$					\\
	$+$				&$0$			&$\varepsilon^2$					\\
	$0$				&$0$			&$\varepsilon$					\\
	$-$				&$0$			&$\varepsilon^2$					\\
	$+$				&$-$			&$\varepsilon$					\\
	$0$				&$-$			&$\varepsilon^2$					\\
	$-$				&$-$			&$0$
\end{tabular}
\end{center}
\caption{\footnotesize{Scaling of the tree-level polarized $WW\rightarrow WW$ amplitudes with the parameter $\varepsilon=m/E$, for $\varepsilon\ll 1$, see text. The Higgs mass $m_H$ is of order $m$. The missing combinations can be obtained by exploiting the $C$ and $P$ symmetry of the $W$ Lagrangian.
The polarizations used in this computation are the $\hat{\epsilon}(p,k)$ defined in (\ref{explpm}); using instead the basis (\ref{helicity}) one finds that the polarized amplitudes that here vanish become of $O(\varepsilon^3)$ or $O(\varepsilon^4)$, depending on whether a longitudinal polarization is present or not. See also \cite{Borel:2012by}.}}
\label{table:covariant}
\end{table}

It is now interesting to ask which pattern we expect if we use the more conventional covariant polarizations (\ref{helicity}).
This computation is performed in \cite{Borel:2012by}, and the result is similar to Table \ref{table:covariant} with the only difference that the polarized amplitudes that here vanish become of $O(\varepsilon^3)$ or $O(\varepsilon^4)$, depending on whether a longitudinal polarization is present or not.
This ``double suppression'' was noticed  in \cite{Borel:2012by}, without explanation.
It is clear that, on the contrary, passing through the axial gauge and our ``axial-covariant'' polarizations we have all the ingredients that are needed to explain this result.
Consider in fact a polarized amplitude computed in covariant gauge using the polarization vectors $\epsilon(p)$ defined in (\ref{helicity}). The result has the general form:
\begin{equation}
\mathcal{A}_{(a,b;c,d)} = \mathcal{A}^{\mu \nu ; \rho \sigma} \,
\epsilon_{1,\mu}^a \, \epsilon_{2,\nu}^b \, (\epsilon_{3,\rho}^c)^* \, (\epsilon_{4,\sigma}^d)^*
\, ,
\end{equation}
where $a,b,c,d = 0, \pm 1$.
Take now this expression, that is linear in all the polarizations, and perform a change of basis by writing them in terms of the $\hat{\epsilon}(p,k)$. What we have to do is to invert the relations (\ref{explpm}), but for the present purpose it is enough to notice that in the limit $\varepsilon \ll 1$ what happens is that the plus and minus polarizations mix at order $\varepsilon^2$, while the longitudinal polarization mixes with the transverse ones at order $\varepsilon$.
Notice that the high-energy behaviour of $\hat{\epsilon}^{0,+,-}(p,k)$ is the same of that of $\epsilon^{0,+,-}(p)$ respectively.
As a consequence, the various polarized subamplitudes mix with each other, and more precisely $\mathcal{A}_{(a,b;c,d)}$ mixes with $\mathcal{A}_{(e,f;g,h)}$ at order:
\begin{equation}
\varepsilon^{|a-e|+ |b-f| + |c-g| + |d-h|}
\, .
\end{equation}
For example, $\mathcal{A}^{(p,k)}_{(++;--)}$ equals zero, with the superscript indicating the basis of the $\hat{\epsilon}(p,k)$ with $k^2=0$, while $\mathcal{A}^{(p,k)}_{(++;+-)}$ is of $O(\varepsilon^2)$. Since $\mathcal{A}^{(p)}_{(++;--)}$ in the basis of the $\epsilon(p)$ receives contributions from $\mathcal{A}^{(p,k)}_{(++;+-)}$ at order $\varepsilon^2$, we conclude that it must be at least of $O(\varepsilon^4)$, as in fact it is since all the other contributions are of the same order or smaller.
In the same way it is easy to see that all the ``double suppressions'' are understood, while the amplitudes that are ``singly suppressed'' (i.e. by $\varepsilon$ or $\varepsilon^2$) remain of the same order in $\varepsilon$.

\section{Example of phenomenological application: EWA} \label{sec:ewa}

The considerations of the previous Sections are of relatively general interest as they are valid in any theory with a gauge symmetry, unbroken (Section \ref{sec:axial:1}) or spontaneously broken (Sections \ref{sec:axial:2} and \ref{sec:covariant}).
Let us now focus on the electroweak sector of the Standard Model, which is one of the most interesting subjects of study at present  from the point of view of high-energy particle phenomenology.
Indeed, uncovering the mechanism behind the spontaneous breaking of the electroweak gauge symmetry (EWSB) is arguably the main purpose of the CERN LHC.
One of the most direct probes of the dynamics of EWSB is the high-energy scattering of the electroweak vector-bosons ($W$), especially the longitudinally-polarized ones.
Since beams of $W$ bosons do not exist, in order to study their interactions among themselves one has to rely on factorization, namely the fact that in a suitable kinematic regime the short-distance $WW$ interactions can be separated from the well-known interaction of the $W$ with the parton that emits it.
As already said in the Introduction the resulting approximation, known as the effective $W$ approximation (EWA) \cite{Kunszt:1987tk}\cite{Borel:2012by}, is obsolete from the computational point of view since the experimental collaborations have enough computing power to simulate the exact process including radiative corrections. Nevertheless the EWA can still be a useful selection tool to understand which kinematic region is most sensitive to the dynamics of EWSB.

One can consider for instance the general process $qX \rightarrow q' Y$ where $q,q'$ are quarks and $X,Y$ are respectively unspecified initial one-particle state and final state.
Refering to \cite{Borel:2012by} for details, we are interested in studying the interactions between $q$ and $X$ mediated by the $W$ bosons and we speak directly of a quark since we have already factored-out the QCD processes\footnote{Meaning that the cross section that we write for an incoming quark has to be convoluted with the Parton Distribution Function of that quark in the colliding hadron.} that produce the quark from a hadron.
Factorization in this case relies on the existence of a large separation of scales between the virtuality $V^2={m^2-p_W^2}$ of the collinear $W$ emission and the hardness $E$ of the relevant subprocess. In a practical situation, this amounts to require forward jets and large transverse momentum of the $W$ bosons in the final state $Y$.
Denoting the momenta of $q$ and $q'$ as:
\begin{equation}
P_q=(E,0,0,E)
\quad , \quad
P_{q'}=(\sqrt{(1-x)^2E^2 + p_\perp^2}, \, p_\perp \cos\phi,\,  p_\perp \sin\phi,\, E(1-x))
\end{equation}
the EWA amounts to say that in the limit $m/E\ll 1$ and $p_\perp/E\ll 1$ the cross section for $qX\rightarrow q' Y$ integrated over $\phi$ can be written as:
\begin{equation} \label{eq:EWA}
\frac{d\sigma(qX\rightarrow q' Y)}{dx dp_\perp} =
\sum_{a=0,\pm1} \frac{C_Q}{2\pi^2} \, f_a(x,p_\perp) \, d\sigma (W^a_Q X \rightarrow Y)
\end{equation}
where $a$ denotes the polarization, $Q$ stands for the charge, and $C_Q$ is a constant that depends on~$Q$.
The $f_a(x,p_\perp)$ are computable splitting functions that describe the collinear emission of the equivalent $W^a$ whose momentum, that should be $P_q - P_{q'}$, is approximated to be on-shell in $d\sigma (W^a_Q X \rightarrow Y)$.

Refering to \cite{Borel:2012by} for discussions and generalizations, let us focus on the case in which $X$ is an incoming $W$ boson while $Y$ consists of two $W$'s.
The proof of (\ref{eq:EWA}) is based on a power expansion of the amplitude of the $WW\rightarrow WW$ subprocess for small $V^2 \sim \mbox{max}\{ m^2, p_\perp^2 \}\ll E^2$:
\begin{equation} \label{eq:expansionEWA}
\mathcal{A}(V^2) = \mathcal{A}(0) + V^2 \mathcal{A}'(0) + O(V^4)
\end{equation}
and on the assumption:
\begin{equation} \label{eq:assumptionEWA}
\mathcal{A}'(0) \sim  \mathcal{A}(0) / E^2.
\end{equation}
This subamplitude represents the hard part of the process, the soft part being the collinear $W$ emission from the quark that determines the splitting functions  $f_a(x,p_\perp)$ in (\ref{eq:EWA}).
Concerning (\ref{eq:assumptionEWA}), two important observations are in order. First of all one has to notice that, while  $\mathcal{A}(0)$ is a gauge invariant quantity, the off-shell amplitude $\mathcal{A}(V^2)$ and its derivative $\mathcal{A}'(0)$ are not. The reason is that, in order to obtain a gauge-invariant result, it is necessary to include the contributions from the ``radiation diagrams'' in which one of the final $W$'s is emitted directly from the quark line. 
As a consequence, $\mathcal{A}'(0)$ can be anything and actually in covariant gauges the condition (\ref{eq:assumptionEWA}) is badly violated, apparently putting in doubt the validity of the EWA.
In fact, although the diagrams that are neglected are not enhanced by any nearly-on-shell propagator, it can happen that in a ``non-physical'' gauge they give results that are much larger than the total, with large cancellations in the sum.
The important point however is that, to prove the validity of the EWA, it is enough to find {\it one} gauge in which (\ref{eq:assumptionEWA}) is valid, and it is shown in \cite{Borel:2012by} that the axial gauge does the job. The EWA is of course true in any gauge, but in some cases its proof can be more subtle.

Another important issue with (\ref{eq:expansionEWA}) is that, as discussed in the previous Sections, there are cases in which the on-shell amplitude $\mathcal{A}(0)$ is suppressed by powers of $\varepsilon=m/E$ in some helicity configurations. Since $V^2$ is typically of order of the squared vector-boson mass $m^2$, it may seem that the EWA (\ref{eq:EWA}) can fail in some cases for the polarized process. In fact if the on-shell amplitude $\mathcal{A}(0)$ times splitting function were suppressed by more powers of $\varepsilon$ than the full amplitude, then one would have to conclude that the dominant contribution comes from the second term in (\ref{eq:expansionEWA}). The EWA seems thus to be violated unless one is able to show that the full amplitude is always as suppressed as the on-shell WW amplitude times splitting function.

Here is an example of a situation in which the machinery of Sections \ref{sec:axial:1}-\ref{sec:covariant} can be of use.
In fact, besides a clear and simple understanding of the pattern of the $\varepsilon$-suppression of the helicity amplitudes involving only vector bosons, we showed how one can easily understand also the suppression of the amplitude involving two fermions and gauge bosons with same helicity.
To prove the validity of the EWA in a channel in which there is some amount of $\varepsilon$-suppression for all the polarizations of the equivalent $W$, such as $q (W)^- \rightarrow q' (W)^+ (W)^+$, it is then sufficient to show that the {\it full} amplitude, including the femions, is suppressed by the same power of $\varepsilon=m/E$ as the dominant $WW\rightarrow WW$ subamplitude.
Since this suppression is independent of $V^2$, it affects the entire $\mathcal{A}(V^2)$ in \ref{eq:expansionEWA} and thus also $\mathcal{A}'(0)$.

Consider the above example of $q (W)^- \rightarrow q' (W)^+ (W)^+$  with massless quarks and the corresponding effective-$W$ process $(W_{eff})^a (W)^- \rightarrow (W)^+ (W)^+$.
From the previous Sections we know that, in the axial gauge with light-like reference vector, the amplitude for the full process vanishes.
On the other hand, looking at the subamplitude times splitting amplitude, when $a=-$ the subamplitude vanishes while when $a=0,+$ it is the splitting amplitude that can be made to vanish by suitably choosing the reference vector as the 4-momentum of one of the two fermions.
Notice in fact that, considering for example the splitting amplitude with a left-handed fermion $\overline{\psi}_-(p_1) {\not \epsilon}^a(k,p) \psi_-(p_2)$, one can make the case $a=-$ vanish by choosing $p=p_1$, or alternatively the case $a=+$ vanish by choosing $p=p_2$  (see  (\ref{eq:fermions:mainpoint})), not both simultaneously. The amplitude with $a=0$ instead vanishes in both cases. 

The situation becomes more involved if one employs the usual covariant-gauge definition of the polarization vectors, denoted by $\epsilon(p)$ in Section \ref{sec:covariant}.
In this case we know that the scaling of the amplitude $\mathcal{A}^a$ of the process $(W_{eff})^a (W)^- \rightarrow (W)^+ (W)^+$ is:
\begin{equation}
\mathcal{A}^+ \sim \varepsilon^2
\quad , \quad
\mathcal{A}^0 \sim \varepsilon^3
\quad , \quad
\mathcal{A}^- \sim \varepsilon^4 \,
\end{equation}
and in the splitting amplitudes there is an additional $\varepsilon$ suppression in the case $a=0$.
The validity of (\ref{eq:EWA}) is not obvious now, unless one is able to show that the full tree-level amplitude including the fermions is suppressed at high energy at least by a factor $\varepsilon^2$.
Without such a proof, in \cite{Borel:2012by} it was checked numerically that in this and similar situations the EWA remains nevertheless valid, as always up to corrections of order $\mbox{max}\{ m^2, p_\perp^2 \}/E^2$.
Using our results it is immediate to understand why the EWA holds: the full amplitude for $q (W)^- \rightarrow q' (W)^+ (W)^+$ is zero in axial gauge, while it is unsuppressed if we flip the helicity of one of the $W$'s. On the other hand we know by (\ref{explpm}) that the polarizations $\epsilon^\pm(p)$ in covariant gauge have a component of $\epsilon^\mp(p,k)$ of order $\varepsilon^2$. As a consequence, the full amplitude is suppressed by $\varepsilon^2$ for any value of the virtuality, and it matches the behaviour of the EWA result.
To be precise, one may question that in covariant gauge eq. (\ref{eq:assumptionEWA}) does not hold in general, so that the second term in (\ref{eq:expansionEWA}) could be of the same order as the first one. However it can be shown that, by suitably modifying the gauge, one can make (\ref{eq:assumptionEWA})  valid also in a certain class of covariant gauges\footnote{This alternative proof of the EWA will be presented in \cite{smartgauge}.}.
Similar situations can be understood analogously, sometimes making use of considerations like those made in in Section \ref{sec:covariant} about the case in which some diagrams are not present because of the conservation of the electric charge.

\section{Conclusions} \label{sec:conclusions}

In this paper we studied some properties of helicity amplitudes in spontaneously broken gauge theories at high energy.
The concept of helicity becomes frame-dependent in the case of massive gauge bosons with mass $m$, and in general it can be ``violated'' by a larger amount with respect to the massless case.
Our main result is the generalization to the massive case of the vanishing theorems that are valid in the case of unbroken gauge theories.
In particular we provided a set of ``selection rules'' that specify by which power of $m/E$ a given helicity amplitude is suppressed at energies of order $E\gg m$.
Moreover we did it using only elementary means, and we also presented a simple discussion of the known results that are valid in the unbroken case.
To show the usefulness of our approach, we discussed as an example how one can employ our methods to understand some aspects of the effective $W$ approximation in the polarized case.

\section*{Acknowledgments}

We thank Roberto Franceschini, Riccardo Rattazzi and Andrea Wulzer for useful discussions. This research is supported by the Swiss National Science Foundation under contract 200021-125237.
We thank the Galileo Galilei Institute for Theoretical Physics for hospitality and the INFN for partial support during the completion of this work.

\appendix

\section{Axial gauge} \label{app:axialgauge}

Refering to \cite{Kunszt:1987tk} and references therein for more details, the axial gauge is defined by the gauge-fixing condition:
\begin{equation}
k_\mu A^\mu =0 \, ,
\end{equation}
that is enforced by a delta function in the functional integration.
An important feature of this choice is that, unlike in the covariant $R_\xi$ gauges, the mixings among Goldstones and gauge fields do not cancel out. As a consequence the propagator is nondiagonal when regarded as a matrix in the 5-dimensional space $(A_\mu,s)$ spanned by the four components of $A_\mu$ plus the Goldstone boson $s$.
Its explicit form is:
\begin{eqnarray}
\mathcal{P}_{IJ}(q,k) &=& \frac{i\, \mathcal{N}_{IJ}(q,k)}{q^2 - m^2}\, , \label{eq:axial:propagator} \\
\mathcal{N}_{\mu \nu}(q,k) &=& -g_{\mu\nu} + \frac{q_\mu k_\nu + q_\nu k_\mu}{qk} - \frac{k^2}{(qk)^2} q_\mu q_\nu \, ,  \nonumber \\
\mathcal{N}_{\mu s}(q,k) &=& \mathcal{N}_{s \mu}(q,k)^* =   -i m \frac{(qk)k_\mu - k^2 q_\mu}{(qk)^2} \, , \nonumber \\
\mathcal{N}_{ss}(q,k) &=& 1-\frac{m^2 k^2}{(qk)^2}   \, . \nonumber
\end{eqnarray}
By construction, the projector matrix $\mathcal{N}_{IJ}$ annihilates the vector $(k^\mu ,0)$ and has thus rank 4 for general $q$. At the pole $q^2 = m^2$, the rank is further reduced down to 3, and the projector can be rewritten as:
\begin{equation}
\label{eq:axial:compl}
\mathcal{N}_{IJ}(q^2=m^2,k) = \sum_{\lambda=+,-,0} \epsilon^\lambda_I(q,k) \epsilon^\lambda_J(q,k)^*
\end{equation}
where the polarization vectors $\epsilon^\pm_I(p,k)$ have vanishing fifth component and satisfy (\ref{eq:axialpol:1})-(\ref{eq:axialpol:3}) while, for $\lambda=0$, the 4-vector $\epsilon^0_I(p,k)$ describes the longitudinal polarization:
\begin{eqnarray}
\epsilon_\mu^0(p,k) &=& \frac{-\frac{m}{p k}k_\mu + \frac{k^2 m}{(p k)^2} p_\mu}
{\sqrt{1-\frac{k^2 m^2}{(p k)^2}}}
\, , \label{eq:axial:longitud} \\
\epsilon_s^0(p,k) &=& -i \sqrt{1-\frac{m^2 k^2}{(pk)^2} } \, .
\nonumber
\end{eqnarray}


Let us consider the transformation properties of these polarization vectors under rotations.
In order to be more concrete, we adopt a ``constructive'' approach.
First of all, from the definition (\ref{eq:axial:longitud}) it is clear that the longitudinal one transforms as:
\begin{equation} \label{eq:trasfpol:long}
\epsilon_\mu^0(Rp,Rk) = R \epsilon_\mu^0(p,k)
\end{equation}
where $R$ is any rotation matrix.
For the transverse states we are particularly interested in rotations along the axes specified by $\vec{p}$ when $p^2=0$ (for any $k^2$), and rotations along $\vec{k}$ with $k^2=0$ (for any $p^2$). 
In general however eq. (\ref{eq:rotationOfPolariz}) holds.

Let us start with a rotation $R_\theta^{\vec{p}}$ of an angle $\theta$ along the particle momentum $\vec{p}$.
Without loss of generality, we can take $\vec{p}$ to be along the third axis and  moreover, since $R_\theta^{\vec{p}}$ commutes with a boost along the same axis, we can go to the frame in which $\vec{k}$ has no third component. The two 4-vectors transform then as:
\begin{eqnarray}
R_\theta^{\vec{p}} p &=& R_\theta^{\vec{p}} (p_0,0,0,|\vec{p}|) =  p \, ,
\\
R_\theta^{\vec{p}} k &=& R_\theta^{\vec{p}}(k_0,k_1,k_2,0) = (k_0,k_1\cos\theta + k_2 \sin\theta , k_2\cos\theta - k_1\sin\theta,0) \, .
\end{eqnarray}
To construct $\epsilon^\pm(p,k)$ we need to choose two 4-vectors $a$ and $b$ with $a^2=b^2=-1/2$ such that $\{a,b,k\}$ form an orthogonal basis of the 3-dimensional subspace orthogonal to $p$, and such that $(\vec{a} \times \vec{b})\cdot\vec{p} >0$ to ensure right-handedness.
In our case we can define\footnote{Notice that in this frame $k_0\neq 0$, because by assumption $p\cdot k\neq 0$.} (for $p^2=0$, i.e. $p_0=|\vec{p}|$):
\begin{equation} \label{eq:aeb}
a(k)=\frac{1}{\sqrt{2}}(\frac{k_1}{k_0},1,0,\frac{k_1}{k_0})
\quad , \quad
b(k)=\frac{1}{\sqrt{2}}(\frac{k_2}{k_0},0,1,\frac{k_2}{k_0}) \, .
\end{equation}
Up to an overall phase factor we can define the transverse polarization vectors  satisfying (\ref{eq:axialpol:1})-(\ref{eq:axialpol:3})  as $\epsilon^\pm(p,k) = a(k) \pm i b(k)$. 
It is then immediate to verify that:
\begin{equation}
\label{ab0}
\epsilon^\pm (R_\theta^{\vec{p}} p, R_\theta^{\vec{p}} k)=
 a(R_\theta^{\vec{p}} k) \pm i b(R_\theta^{\vec{p}} k) =
e^{\pm i\theta} R_\theta^{\vec{p}} (a(k) \pm ib(k))=
e^{\pm i\theta} R_\theta^{\vec{p}} \epsilon^\pm (p, k)
\qquad [p^2=0]\, .
\end{equation}

Consider now the case of a rotation along the direction of the reference vector $\vec{k}$ with $k^2=0$.
Everything proceeds as above, with the roles of $p$ and $k$ interchanged, except for the choice of $a$ and $b$.
In fact, with the definition (\ref{eq:aeb}), one has $(\vec{a} \times \vec{b})=(-k_1/k_0, -k_2/k_0,1)$ so that $(\vec{a}(k) \times \vec{b}(k))\cdot\vec{p} >0$ but when the construction is done with $k\leftrightarrow p$ one has to choose the opposite ordering for $a(p)$ and $b(p)$.
As a consequence:
\begin{equation} 
\epsilon^\pm (R_\theta^{\vec{k}} p, R_\theta^{\vec{k}} k)=
 b(R_\theta^{\vec{k}} p) \pm i a(R_\theta^{\vec{k}} p) =
e^{\mp i\theta} R_\theta^{\vec{k}} (b(p) \pm ia(p))=
e^{\mp i\theta} R_\theta^{\vec{k}} \epsilon^\pm (p, k)
\qquad [k^2=0]\, .
\end{equation}

To consider the more general cases in which we rotate for example along $\vec{p}$ with $p^2 = m^2 > 0$, one can notice that the expression for $a(k)$ and $b(k)$ in (\ref{eq:aeb}) can be generalized to:
\begin{equation}
\begin{split}
a(k,p)&=\frac{1}{\sqrt{2}}\left( \beta_p \frac{k_1}{k_0}, 1- \frac{k_1^2(1-\beta_p)}{k_0^2}, - \frac{k_1 k_2 (1-\beta_p)}{k_0^2} , \frac{k_1}{k_0} \right)\\
b(k,p)&=\frac{1}{\sqrt{2}}\left( \beta_p \frac{k_2}{k_0}, - \frac{k_1 k_2 (1-\beta_p)}{k_0^2}, 1- \frac{k_2^2(1-\beta_p)}{k_0^2}, \frac{k_2}{k_0} \right) \, ,
\end{split}
\end{equation}
where $\beta_p = |\vec{p}|/\sqrt{|\vec{p}|^2 + m^2}$, and the same transformation rules still apply.

To summarize, the relevant transformation properties of the transverse polarization vectors can be written as:
\begin{eqnarray} 
R_\theta^{\vec{p}} \epsilon^\pm (p, k) &=&
e^{\mp i\theta}  \epsilon^\pm (R_\theta^{\vec{p}} p=p, R_\theta^{\vec{p}} k)
\, , \label{eq:trasfpol:transv1} \\
R_\theta^{\vec{k}} \epsilon^\pm (p, k) &=&
e^{\pm i\theta} \epsilon^\pm (R_\theta^{\vec{k}} p, R_\theta^{\vec{k}} k=k)
\, . \label{eq:trasfpol:transv2}
\end{eqnarray}
Notice that the fact that the rotation along the $\vec{k}$ axis brings a phase which is opposite to that from a rotation along the $\vec{p}$ axis is what one expects from (\ref{eq1:massless}) and (\ref{eq2:massless}).

\vspace{0.3cm}



\end{document}